\documentclass[prd,nopacs,preprintnumbers,twocolumn,amsmath,nofootinbib,amssymb]{revtex4}
\usepackage{graphicx,color,dcolumn,booktabs,bm}
\usepackage{longtable,lscape}
\usepackage{txfonts}
\usepackage{overpic}
\usepackage{amssymb}
\usepackage{epstopdf}
\usepackage{indentfirst}
\usepackage{feynmf}   
\usepackage{slashed}  
\usepackage{cases}
\usepackage{color}
\usepackage{float}
\usepackage{multirow}
\usepackage{makecell}
\usepackage{graphicx,color,dcolumn,booktabs,bm}
\usepackage{epsfig,dsfont,amssymb,amsmath,amsfonts,amsbsy,mathrsfs}
\usepackage{ulem}

\graphicspath{{Figures/}} %

\usepackage{hyperref}
\hypersetup{colorlinks,citecolor=blue,anchorcolor=red,menucolor=red, linkcolor=red,filecolor=red,runcolor=red,urlcolor=blue,frenchlinks=red}

\makeatletter
\@addtoreset{equation}{section}
\makeatother

\allowdisplaybreaks

\begin{document}

\title{Newly observed $X(4630)$: a new charmoniumlike molecule}

\author{Xin-Dian Yang$^{1,2}$}
\email{yangxd20@lzu.edu.cn}
\author{Fu-Lai Wang$^{1,2}$}
\email{wangfl2016@lzu.edu.cn}
\author{Zhan-Wei Liu$^{1,2,3}$}\email{liuzhanwei@lzu.edu.cn}
\author{Xiang Liu$^{1,2,3}$}
\email{xiangliu@lzu.edu.cn}
\affiliation{$^1$School of Physical Science and Technology, Lanzhou University, Lanzhou 730000, China\\
$^2$Research Center for Hadron and CSR Physics, Lanzhou University and Institute of Modern Physics of CAS, Lanzhou 730000, China\\
$^3$Lanzhou Center for Theoretical Physics, Key Laboratory of Theoretical Physics of Gansu Province, and Frontiers Science Center for Rare Isotopes, Lanzhou University, Lanzhou 730000, China}

\begin{abstract}
Very recently, the LHCb Collaboration at the Large Hadron Collider at CERN observed new resonance $X(4630)$. The $X(4630)$ is decoded as a charmoniumlike molecule with hidden-strange quantum number well in the one-boson-exchange mechanism. Especially, the study of its hidden-charmed decays explicitly shows the dominant role of $J/\psi\phi$ among all allowed hidden-charmed decays of the $X(4630)$, which enforces the conclusion of $X(4630)$ as a charmoniumlike molecule. The discovery of the $X(4630)$ is a crucial step of constructing charmoniumlike molecule zoo.
\end{abstract}

\maketitle

\section{Introduction}
In 1964, Gell-Mann \cite{GellMann:1964nj} and Zweig \cite{Zweig:1981pd} independently proposed a scheme to categorize one hundred hadrons by introducing SU(3) symmetry. However, at the birth of quark model, they had never imagined that the situation of the observation of hadronic states would become so prosperous. In the past two decades, experiment has made big achievement of observation of new hadrons. Especially, LHC experiments, as a frontier of exploring micro-structure of matter, have found more than 50 new hadrons over the past 10 years, which as a big news was released by LHC  \cite{lhcb}.
As the theory describing how the strong interaction makes quarks to be bound together for forming different hadrons, quantum chromodynamics (QCD) has peculiar non-perturbative behavior which has close relation to colour confinement.
How to mathematically depict it is full of challenges. These observations of new hadrons are promoting our understanding to non-perturbative behavior of strong interaction \cite{Chen:2016qju,Liu:2019zoy,Olsen:2017bmm,Guo:2017jvc,Liu:2013waa,Hosaka:2016pey,Brambilla:2019esw}.

On March 3 2021, the LHCb Collaboration announced new measurement of new resonances in the $J/\psi K^+$ and $J/\psi\phi$ invariant mass distribution of $B^+\to J/\psi \phi K^+$ \cite{Aaij:2021ivw}. Among these newly reported states, the $X(4630)$ as a new state was discovered in the $J/\psi\phi$ invariant mass distribution, which has resonance parameter
\begin{eqnarray}
M&=&4626\pm 16^{+18}_{-110}\,{\rm MeV},\nonumber\\
\Gamma&=&174\pm27^{+134}_{-73}\,\rm MeV.\nonumber
\end{eqnarray}
In addition, an important information of its spin-parity quantum number was measured, where $J^P$ of the $X(4630)$ favors $1^-$ \cite{Aaij:2021ivw}.

This novel observation of the $X(4630)$ inspires us to decode the $X(4630)$. Due to the constraint from $C$ parity conservation, the $C$ parity of the $X(4630)$ must be positive, which is deduced from its decay channel $J/\psi\phi$ \cite{Aaij:2021ivw}. Obviously, the $X(4630)$ with $1^{-+}$ must be an absolute exotic state different from conventional meson since $1^{-+}$ is not allowed for conventional meson. We may notice an interesting fact, i.e., the observed $X(4630)$ is just near the $D_s^{*} D_{s1}(2536)$ threshold, which makes us naturally propose the newly observed $X(4630)$ as a $D_s^{*} \bar  D_{s1}(2536)$ charmoniumlike molecule. Here, we need to mention that there were the theoretical investigations of this type hidden-charm hadronic molecular states with hidden strangeness in the past years \cite{Wang:2020dya,Dong:2021juy,Wang:2020cme}.

In this paper, we will give a mass spectrum analysis to reflect the fact of the $X(4630)$ as a new type of charmoniumlike molecule, where the mass of the $X(4630)$ can be reproduced well under a $D_s^{*} \bar D_{s1}(2536)$ charmoniumlike molecule assignment. Coming with a further hidden-charmed decay study, one understands why the $X(4630)$ was firstly observed in its $J/\psi\phi$ since our study shows that $J/\psi\phi$ is a main decay channel among its allowed decays. Our study provides direct evidence of the $X(4630)$ as a $D_s^{*} \bar D_{s1}(2536)$ charmoniumlike molecule. Decoding the $X(4630)$ as charmoniumlike molecule with hidden-strange quantum number is a crucial step in constructing charmoniumlike molecule family. As an extension, we further predict spin partners of the $X(4630)$, which can be new tasks for future experiment at LHC.

\section{Mass spectrum analysis} \label{secMass}
In this paper, we study whether the $X(4630)$ can be identified as a $D_s^{*} \bar D_{s1}(2536)$ charmoniumlike molecule. First, we study the interactions between the charmed-strange meson $D_s^{*}$ and the anticharmed-strange meson $\bar D_{s1}(2536)$ in the one-boson-exchange (OBE) mechanism, which is often adopted to study the heavy flavored hadrons interactions and identify these observed $X/Y/Z/P_c$ states in a hadronic molecular picture \cite{Chen:2016qju,Liu:2019zoy}. Here, we include the contribution from the $f_0(980)$, $\eta$, and $\phi$ exchanges for the $D_s^{*}\bar D_{s1}$ system \cite{Wang:2021hql}.

By considering the heavy quark symmetry, chiral symmetry, and hidden local symmetry \cite{Casalbuoni:1992gi,Casalbuoni:1996pg,Yan:1992gz,Harada:2003jx,Bando:1987br}, the effective Lagrangians for the (anti-)charmed mesons with the light mesons are constructed as
\begin{eqnarray}\label{eq:compactlag}
{\cal L}&=&g_{\sigma}\left\langle H^{(Q)}_a\sigma\overline{H}^{(Q)}_a\right\rangle+g_{\sigma}\left\langle \overline{H}^{(\overline{Q})}_a\sigma H^{(\overline{Q})}_a\right\rangle\nonumber\\
&&+g^{\prime\prime}_{\sigma}\left\langle T^{(Q)\mu}_a\sigma\overline{T}^{(Q)}_{a\mu}\right\rangle+g^{\prime\prime}_{\sigma}\left\langle\overline{T}^{(\overline{Q})\mu}_a\sigma T^{(\overline{Q})}_{a\mu}\right\rangle\nonumber\\
&&+\frac{h^{\prime}_{\sigma}}{f_{\pi}}\left[\left\langle T^{(Q)\mu}_a\partial_{\mu}\sigma\overline{H}^{(Q)}_b\right\rangle+\left\langle\overline{H}^{(\overline{Q})}_a\partial_{\mu}\sigma T^{(\overline{Q})\mu}_b\right\rangle+h.c.\right]\nonumber\\
&&+ig\left\langle H^{(Q)}_b{\cal A}\!\!\!\slash_{ba}\gamma_5\overline{H}^{\,({Q})}_a\right\rangle+ig\left\langle \overline{H}^{(\overline{Q})}_a{\cal A}\!\!\!\slash_{ab}\gamma_5 H^{\,(\overline{Q})}_b\right\rangle\nonumber\\
&&+ik\left\langle T^{\,(Q)\mu}_b{\cal A}\!\!\!\slash_{ba}\gamma_5\overline{T}^{(Q)}_{a\mu}\right\rangle+ik\left\langle\overline{T}^{\,(\overline{Q})\mu}_a{\cal A}\!\!\!\slash_{ab}\gamma_5T^{(\overline{Q})}_{b\mu}\right\rangle\nonumber\\
&&+\left[i\left\langle T^{(Q)\mu}_b\left(\frac{h_1}{\Lambda_{\chi}}D_{\mu}{\cal A}\!\!\!\slash+\frac{h_2}{\Lambda_{\chi}}D\!\!\!\!/ {\cal A}_{\mu}\right)_{ba}\gamma_5\overline{H}^{\,(Q)}_a\right\rangle+H.c.\right]\nonumber\\
&&+\left[i\left\langle\overline{H}^{\,(\overline{Q})}_a\left(\frac{h_1}{\Lambda_{\chi}}{\cal A}\!\!\!\slash\stackrel{\leftarrow}{D_{\mu}'}+\frac{h_2}{\Lambda_{\chi}}{\cal A}_{\mu}\stackrel{\leftarrow}{D\!\!\!\slash'}\right)_{ab}\gamma_5T^{(\overline{Q})\mu}_b\right\rangle+H.c.\right]\nonumber\\
&&+\left\langle iH^{(Q)}_b\left(\beta v^{\mu}({\cal V}_{\mu}-\rho_{\mu})+\lambda \sigma^{\mu\nu}F_{\mu\nu}(\rho)\right)_{ba}\overline{H}^{\,(Q)}_a\right\rangle\nonumber\\
&&-\left\langle i\overline{H}^{(\overline{Q})}_a\left(\beta v^{\mu}({\cal V}_{\mu}-\rho_{\mu})-\lambda \sigma^{\mu\nu}F_{\mu\nu}(\rho)\right)_{ab}H^{\,(\overline{Q})}_b\right\rangle\nonumber\\
&&+\left\langle iT^{\,(Q)}_{b\lambda}\left(\beta^{\prime\prime} v^{\mu}({\cal V}_{\mu}-\rho_{\mu})+\lambda^{\prime\prime}\sigma^{\mu\nu}F_{\mu\nu}(\rho)\right)_{ba}\overline{T}^{(Q)\lambda}_{a}\right\rangle\nonumber\\
&&-\left\langle i\overline{T}^{\,(\overline{Q})}_{a\lambda}\left(\beta^{\prime\prime} v^{\mu}({\cal V}_{\mu}-\rho_{\mu})-\lambda^{\prime\prime}\sigma^{\mu\nu}F_{\mu\nu}(\rho)\right)_{ab}T^{(\overline{Q})\lambda}_{b}\right\rangle\nonumber\\
&&+\left[\left\langle T^{(Q)\mu}_b\left(i\zeta_1({\cal V}_{\mu}-\rho_{\mu})+\mu_{1}\gamma^{\nu}F_{\mu\nu}(\rho)\right)_{ba}\overline{H}^{\,(Q)}_a\right\rangle+H.c.\right]\nonumber\\
&&-\left[\left\langle\overline{H}^{\,(\overline{Q})}_a\left(i\zeta_1({\cal V}_{\mu}-\rho_{\mu})-\mu_1\gamma^{\nu}F_{\mu\nu}(\rho)\right)_{ab}T^{(\overline{Q})\mu}_b\right\rangle+H.c.\right],\nonumber\\
\end{eqnarray}
where the covariant derivatives  can be written as $D_{\mu}=\partial_{\mu}+{\cal V}_{\mu}$ and $D'_{\mu}=\partial_{\mu}-{\cal V}_{\mu}$, and the vector meson field $\rho_{\mu}$ and its strength tensor $F_{\mu\nu}(\rho)$ defined as $\rho_{\mu}=i{g_V}\mathbb{V}_{\mu}/{\sqrt{2}}$ and $F_{\mu\nu}(\rho)=\partial_{\mu}\rho_{\nu}-\partial_{\nu}\rho_{\mu}+[\rho_{\mu},\rho_{\nu}]$, respectively. In the above expressions, the vector current ${\cal V}_{\mu}$ and the axial current $\mathcal{A}_\mu$ are
\begin{eqnarray}
{\mathcal V}_{\mu}&=&\frac{1}{2}(\xi^{\dagger}\partial_{\mu}\xi+\xi\partial_{\mu}\xi^{\dagger}),\\
{\mathcal A}_{\mu}&=&\frac{1}{2}(\xi^\dag\partial_\mu\xi-\xi \partial_\mu\xi^\dag),
\end{eqnarray}
with $\xi=\exp(i\mathbb{P}/f_\pi)$ and the pion decay constant is taken as $f_\pi=132~\rm{MeV}$. The light pseudoscalar meson matrix ${\mathbb{P}}$ and the light vector meson matrix $\mathbb{V}_{\mu}$ have the standard form, i.e.,
\begin{eqnarray}
\left.\begin{array}{c}
{\mathbb{P}} = {\left(\begin{array}{ccc}
       \frac{\pi^0}{\sqrt{2}}+\frac{\eta}{\sqrt{6}} &\pi^+ &K^+\\
       \pi^-       &-\frac{\pi^0}{\sqrt{2}}+\frac{\eta}{\sqrt{6}} &K^0\\
       K^-         &\bar K^0   &-\sqrt{\frac{2}{3}} \eta     \end{array}\right)},\\
{\mathbb{V}}_{\mu} = {\left(\begin{array}{ccc}
       \frac{\rho^0}{\sqrt{2}}+\frac{\omega}{\sqrt{2}} &\rho^+ &K^{*+}\\
       \rho^-       &-\frac{\rho^0}{\sqrt{2}}+\frac{\omega}{\sqrt{2}} &K^{*0}\\
       K^{*-}         &\bar K^{*0}   & \phi     \end{array}\right)}_{\mu}.
\end{array}\right.
\end{eqnarray}
The superfields relating to the (anti-)charmed mesons can be defined by \cite{Ding:2008gr}
\begin{eqnarray}
H^{(Q)}_a&=&{\cal P}_{+}\left(D^{*(Q)\mu}_a\gamma_{\mu}-D^{(Q)}_a\gamma_5\right),\nonumber\\
T^{(Q)\mu}_a&=&{\cal P}_{+}\left[D^{*(Q)\mu\nu}_{2a}\gamma_{\nu}-\sqrt{\frac{3}{2}}D^{(Q)}_{1a\nu}\gamma_5\left(g^{\mu\nu}-\frac{1}{3}\gamma^{\nu}\left(\gamma^{\mu}-v^{\mu}\right)\right)\right],\nonumber\\
H^{(\overline{Q})}_a&=&\left(\bar{D}^{*(\overline{Q})\mu}_{a}\gamma_{\mu}-\bar{D}^{(\overline{Q})}_a\gamma_5\right){\cal P}_{-},\nonumber\\
T^{(\overline{Q})\mu}_{a}&=&\left[\bar{D}^{*(\overline{Q})\mu\nu}_{2a}\gamma_{\nu}-\sqrt{\frac{3}{2}}\bar{D}^{(\overline{Q})}_{1a\nu}\gamma_5\left(g^{\mu\nu}-\frac{1}{3}\left(\gamma^{\mu}-v^{\mu}\right)\gamma^{\nu}\right)\right]{\cal P}_{-},\nonumber\\
\end{eqnarray}
where the projection operator ${\cal P}_{\pm}=(1\pm{v}\!\!\!\slash)/2$ and the velocity $v^{\mu}=(1,\,\bf {0})$. In addition, their conjugate fields satisfy
$\overline{X}=\gamma_0X^{\dagger}\gamma_0$ with $X=H^{(Q)}_a,\,T^{(Q)\mu}_a,\,H^{(\overline{Q})}_a,\,T^{(\overline{Q})\mu}_{a}$. By expanding the compact effective Lagrangians to the leading order of the pseudo-Goldstone field, the detailed effective Lagrangians for the (anti-)charmed mesons and the exchanged light mesons can be obtained.

With above effective Lagrangians, we further write out the scattering amplitudes $\mathcal{M}(h_1h_2\to h_3h_4)$ of the scattering process $h_1h_2\to h_3h_4$ by the effective Lagrangians approach. For the $D_s^{*}\bar D_{s1}$ system, there exist the direct channel and crossed channel Feynman diagrams. The effective potential in momentum space $\mathcal{V}^{h_1h_2\to h_3h_4}(\bm{q})$ can be related to the scattering amplitude $\mathcal{M}(h_1h_2\to h_3h_4)$ \cite{Breit:1929zz,Breit:1930zza}
\begin{eqnarray}\label{breit}
\mathcal{V}_E^{h_1h_2\to h_3h_4}(\bm{q}) &=&-\frac{\mathcal{M}(h_1h_2\to h_3h_4)} {\sqrt{\prod_i2m_i\prod_f2m_f}},
\end{eqnarray}
where $m_i$ and $m_f$ are the masses of the initial states $(h_1, \,h_2)$ and final states $(h_3, \,h_4)$, respectively.  The effective potential in the coordinate space $\mathcal{V}^{h_1h_2\to h_3h_4}(\bm{r})$ can be deduced via the Fourier transformation. In order to  compensate the effects from the off-shell exchanged mesons and more complicate structure of hadrons \cite{Tornqvist:1993ng,Tornqvist:1993vu}, the monopole type form factor $\mathcal{F}(q^2,m_E^2) = (\Lambda^2-m_E^2)/(\Lambda^2-q^2)$ is introduced at every interactive vertex with $m_E$ and $q$ are the mass and four-momentum of the exchanged particle. Here, the cutoff $\Lambda$ is a parameter of the OBE mechanism, and we attempt to find bound state solutions by varying the cutoff parameter in the present work.

In addition, the normalized relations for  the vector charmed-strange meson $D_s^{*}$ and the axial-vector charmed-strange meson $D_{s1}$ satisfy
\begin{eqnarray}
\left.\begin{array}{ll}
\langle 0|D_s^{*\mu}|c\bar{s}(1^-)\rangle=\epsilon^\mu\sqrt{m_{D_s^*}},\quad&\langle 0|D_{s1}^{\mu}|c\bar{s}(1^+)\rangle=\epsilon^\mu\sqrt{m_{D_{s1}}},\\
\end{array}\right.
\end{eqnarray}
respectively. In the above expressions, the explicit expressions for the polarization vector $\epsilon_{m}^{\mu}\,(m=0,\,\pm1)$ with spin-1 field is written as $\epsilon_{\pm}^{\mu}= \left(0,\,\pm1,\,i,\,0\right)/\sqrt{2}$ and $\epsilon_{0}^{\mu}= \left(0,0,0,-1\right)$ in the static limit. And then, the spin-orbital wave functions $|{}^{2S+1}L_{J}\rangle$ for the investigated $D_s^{*}\bar D_{s1}$ molecular system is constructed as
\begin{eqnarray}
|D_s^{*}\bar D_{s1}\rangle &=&\sum_{m,m^{\prime},m_S,m_L}C^{S,m_S}_{1m,1m^{\prime}}C^{J,M}_{Sm_S,Lm_L}\epsilon_{m}^\mu\epsilon_{m^{\prime}}^\nu|Y_{L,m_L}\rangle.
\end{eqnarray}
Here, the constant $C^{e,f}_{ab,cd}$ is the Clebsch-Gordan coefficient, and $|Y_{L,m_L}\rangle$ stands for the spherical harmonics function.

With the above preparation, we give the OBE effective potentials for the $D_s^{*}\bar D_{s1}$ system which are composed of the direct channel potential $\mathcal{V}_{D}$ and the cross channel potential $\mathcal{V}_{C}$
\begin{eqnarray}
\mathcal{V}_{D}&=&g_{\sigma}g_{\sigma}^{\prime\prime}\mathcal{O}_1 Y_{f_0}+\frac{5g k}{27f_\pi^2}\left[\mathcal{O}_2\mathcal{Z}+\mathcal{O}_3\mathcal{T}\right]Y_{\eta}\nonumber\\
&&+\left[\frac{\beta \beta^{\prime\prime} g_{V}^2}{2}\mathcal{O}_1+\frac{5\lambda \lambda^{\prime\prime}g_V^2}{9}\left(\mathcal{O}_3\mathcal{T}-2\mathcal{O}_2\mathcal{Z}\right)\right]Y_{\phi},\\
\mathcal{V}_{C}&=&\frac{h_\sigma^{\prime2}}{18f_{\pi}^2}\left[\mathcal{O}_2\mathcal{Z}+\mathcal{O}_{3}\mathcal{T}\right]Y_{{f_0}0}+\frac{\zeta_1^2g_{V}^2}{12}\mathcal{O}_{2}Y_{\phi0}\nonumber\\
&&+\frac{h^{\prime2}}{9f_\pi^2}\left[\mathcal{O}_{4}\mathcal{Z}\mathcal{Z}+\mathcal{O}_{5}\mathcal{T}\mathcal{T}+\mathcal{O}_{6}\left(\mathcal{T}\mathcal{Z}+\mathcal{Z}\mathcal{T}\right)\right]Y_{\eta0},\nonumber\\
\end{eqnarray}
where the function $Y_{E0}$ reads as
\begin{eqnarray}
Y_{E0}= \dfrac{e^{-m_{E0}r}-e^{-\Lambda_0r}}{4\pi r}-\dfrac{\Lambda_0^2-m_{E0}^2}{8\pi\Lambda_0}e^{-\Lambda_0r}.
\end{eqnarray}
Here, $m_{E0}=\sqrt{m_{E}^2-q_0^2}$ and $\Lambda_0=\sqrt{\Lambda^2-q_0^2}$ with $q_0 = 0.42$ GeV, and we define $\mathcal{Z}=\frac{1}{r^2}\frac{\partial}{\partial r}r^2\frac{\partial}{\partial r}$ and $\mathcal{T}=r\frac{\partial}{\partial r}\frac{1}{r}\frac{\partial}{\partial r}$.
In addition, we introduce several operators, which have the form of
\begin{eqnarray}\label{op}
\mathcal{O}_{1}&=&\left({\bm\epsilon^{\dagger}_3}\cdot{\bm\epsilon_1}\right)\left({\bm\epsilon^{\dagger}_4}\cdot{\bm\epsilon_2}\right),~~~~~~\mathcal{O}_{2}=\left({\bm\epsilon^{\dagger}_3}\times{\bm\epsilon_1}\right)\cdot\left({\bm\epsilon^{\dagger}_4}\times{\bm\epsilon_2}\right),\nonumber\\
\mathcal{O}_{3}&=&S({\bm\epsilon^{\dagger}_3}\times{\bm\epsilon_1},{\bm\epsilon^{\dagger}_4}\times{\bm\epsilon_2},\hat{\bm r}),\nonumber\\
\mathcal{O}_{4}&=&-\frac{1}{3}\left({\bm\epsilon^{\dagger}_3}\cdot{\bm\epsilon_1}\right)\left({\bm\epsilon^{\dagger}_4}\cdot{\bm\epsilon_2}\right)
+\frac{1}{3}\left({\bm\epsilon^{\dagger}_3}\cdot{\bm\epsilon^{\dagger}_4}\right)\left({\bm\epsilon_1}\cdot{\bm\epsilon_2}\right),\nonumber\\
\mathcal{O}_{5}&=&\frac{2}{3}S({\bm\epsilon^{\dagger}_3},{\bm\epsilon_1},\hat{\bm r})S({\bm\epsilon^{\dagger}_4},{\bm\epsilon_2},\hat{\bm r})+\frac{1}{3}S({\bm\epsilon^{\dagger}_3},{\bm\epsilon^{\dagger}_4},\hat{\bm r})S({\bm\epsilon_1},{\bm\epsilon_2},\hat{\bm r}),\nonumber\\
\mathcal{O}_{6}&=&\frac{1}{6}\left({\bm\epsilon^{\dagger}_3}\cdot{\bm\epsilon^{\dagger}_4}\right)S({\bm\epsilon_1},{\bm\epsilon_2},\hat{\bm r})+\frac{1}{6}\left({\bm\epsilon_1}\cdot{\bm\epsilon_2}\right)S({\bm\epsilon^{\dagger}_3},{\bm\epsilon^{\dagger}_4},\hat{\bm r})\nonumber\\
&&-\frac{1}{3}\left({\bm\epsilon^{\dagger}_3}\cdot{\bm\epsilon_1}\right)S({\bm\epsilon^{\dagger}_4},{\bm\epsilon_2},\hat{\bm r}),
\end{eqnarray}
with $S\left({\bm{a}},{\bm{b}},\hat{\bm r}\right)= 3\left(\hat{\bm r} \cdot {\bm a}\right)\left(\hat{\bm r} \cdot {\bm b}\right)-{\bm a} \cdot {\bm b}$, and these relevant operators $\mathcal{O}_{i}[J]$ should be sandwiched between the discussed spin-orbit wave functions in a matrix form, such as $\mathcal{O}_{1}[0]=\rm {diag}(1,1)$, $\mathcal{O}_{2}[0]=\rm {diag}(2,-1)$, $\mathcal{O}_{3}[0]=\left(\begin{array}{cc} 0 & \sqrt{2} \\ \sqrt{2} & 2\end{array}\right)$, $\mathcal{O}_{4}[0]=\rm {diag}(\frac{2}{3},-\frac{1}{3})$, $\mathcal{O}_{5}[0]=\left(\begin{array}{cc} \frac{4}{3} & -\frac{2\sqrt{2}}{3} \\ -\frac{2\sqrt{2}}{3} & 4\end{array}\right)$, $\mathcal{O}_{6}[0]=\left(\begin{array}{cc} 0 & \frac{\sqrt{2}}{15} \\ -\frac{8\sqrt{2}}{15} & -\frac{1}{15}\end{array}\right)$, $\mathcal{O}_{1}[1]=\rm {diag}(1,1,1)$, $\mathcal{O}_{2}[1]=\rm {diag}(1,1,-1)$, $\mathcal{O}_{3}[1]=\left(\begin{array}{ccc} 0 & -\sqrt{2} &0 \\ -\sqrt{2} & 1 &0 \\ 0&0&1\end{array}\right)$, $\mathcal{O}_{4}[1]=\rm {diag}(-\frac{1}{3},-\frac{1}{3},-\frac{1}{3})$, $\mathcal{O}_{5}[1]=\left(\begin{array}{ccc} -\frac{2}{3} & -\frac{2\sqrt{2}}{3} &0 \\ -\frac{2\sqrt{2}}{3} & 0 &0 \\ 0&0&-\frac{4}{3}\end{array}\right)$, $\mathcal{O}_{6}[1]=\left(\begin{array}{ccc} 0 & -\frac{1}{30\sqrt{2}}&\frac{\sqrt{3}}{10\sqrt{2}} \\ -\frac{1}{30\sqrt{2}}&\frac{4}{105} &\frac{3\sqrt{3}}{70} \\ \frac{\sqrt{3}}{10\sqrt{2}}&\frac{3\sqrt{3}}{70}&-\frac{1}{105}\end{array}\right)$, and so on \cite{Wang:2020dya}.
In the following numerical analysis, the coupling constants are $g_\sigma=0.76$,  $g_\sigma^{\prime\prime}=-0.76$, $h_{\sigma}^{\prime}=0.35$, $g=0.59$, $k=0.59$, $|h^{\prime}|=0.55~\rm{GeV}^{-1}$, $f_\pi=0.132~\rm{GeV}$, $\beta=-0.90$, $\beta^{\prime\prime}=0.90$, $\lambda=-0.56~\rm{GeV}^{-1}$, $\lambda^{\prime\prime}=0.56~\rm{GeV}^{-1}$, $|\zeta_1|=0.20$, $\mu_1=0$, and $g_V=5.83$ \cite{Wang:2020dya,Casalbuoni:1996pg,Falk:1992cx,Isola:2003fh,Cleven:2016qbn,Dong:2019ofp,He:2019csk,Wang:2019nwt,Wang:2019aoc,Wang:2020lua,Riska:2000gd}, and the adopted hadron masses are $m_{f_0}=990.00~\rm{MeV}$, $m_{\eta} =547.86~\rm{MeV}$, $m_{\phi}=1019.46~\rm{MeV}$, $m_{D_s^{\ast}}=2112.20~\rm{MeV}$, and $m_{D_{s1}(2536)}=2535.11~\rm{MeV}$ \cite{Zyla:2020zbs}.

For the $D_s^{*}\bar D_{s1}$ molecular system, we need distinguish the charge parity quantum numbers $C$ due to the charge conjugate transformation invariance, and the flavor wave function $|I,I_{3}\rangle$ is defined as $\left|0,0\right\rangle =\left|D_s^{*+} D_{s1}^{-}+cD_{s1}^{+} D_s^{*-}\right\rangle/\sqrt{2}$, where $C=-c\cdot(-1)^{2-J}$ with $J$ is the total spin of the $D_s^{*}\bar D_{s1}$ system \cite{Wang:2020dya,Liu:2008fh,Liu:2008tn,Sun:2012sy}.

\begin{figure}[!tbp]
\includegraphics[width=4.20cm,keepaspectratio]{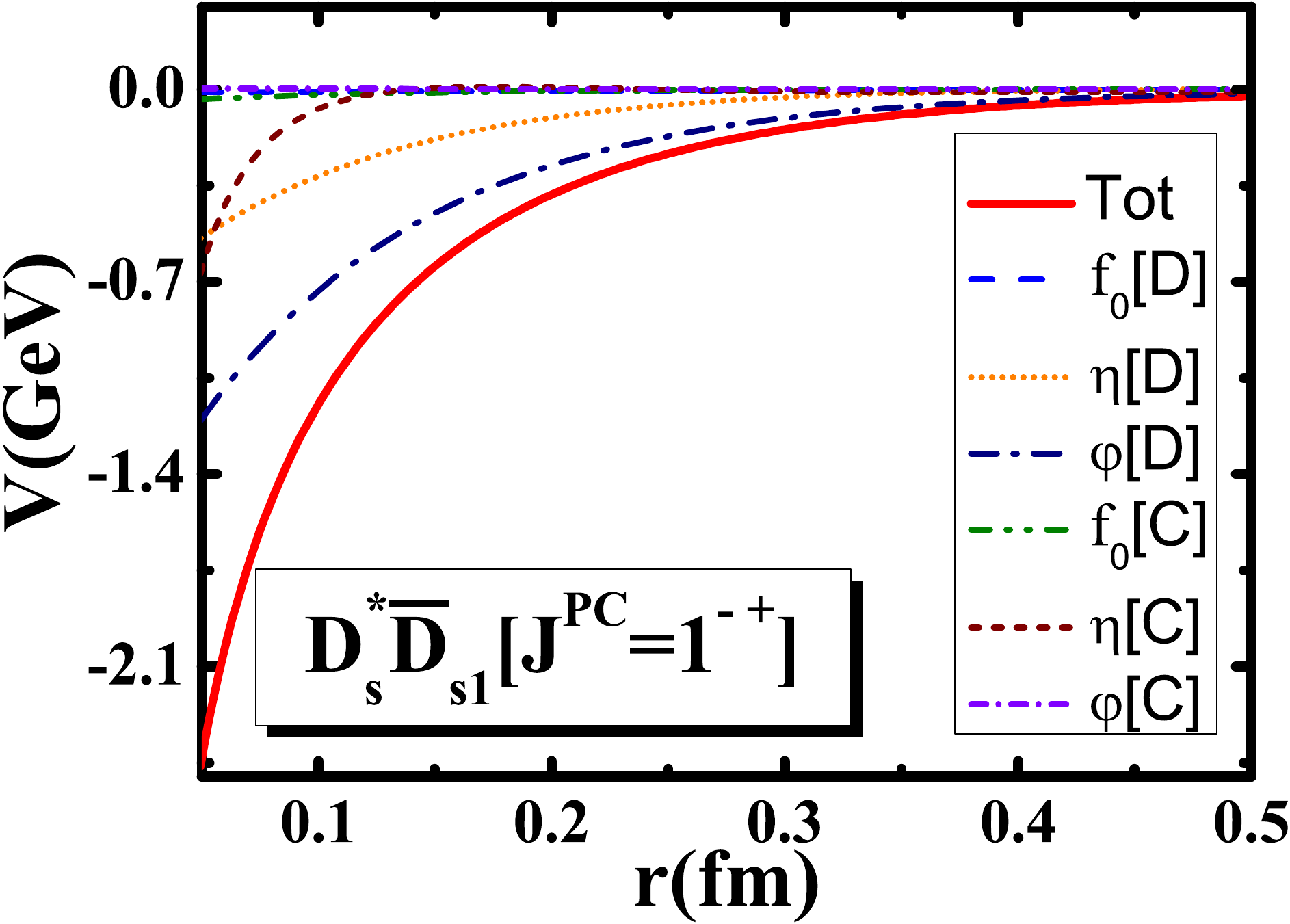}~\includegraphics[width=4.20cm,keepaspectratio]{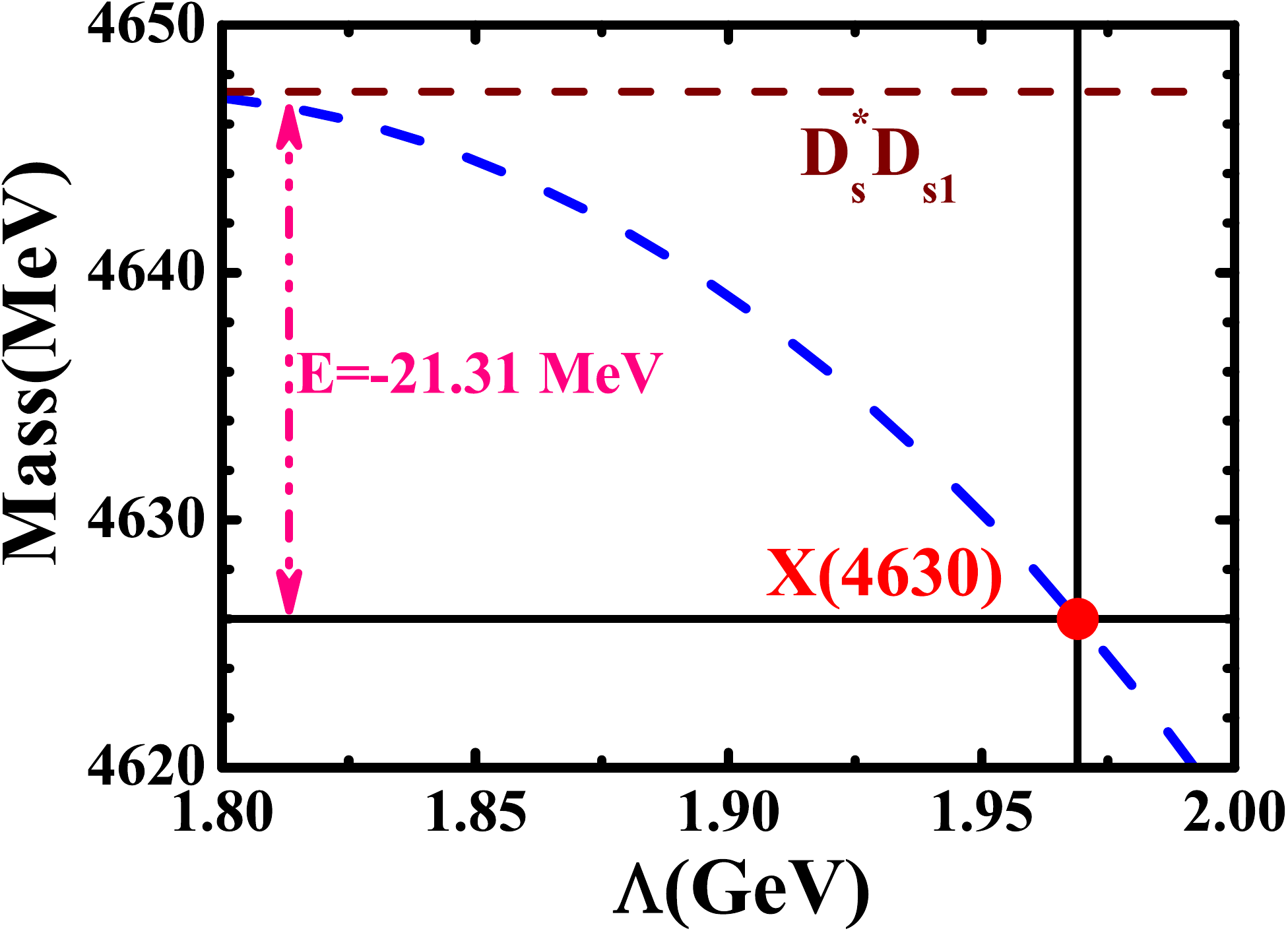}
\caption{The bound state solutions for the $D_s^{*}\bar D_{s1}$ state with $J^{PC}=1^{-+}$. The left diagram is the $r$ dependence of the OBE effective potentials with the cutoff $\Lambda=1.97$ GeV, and the right diagram is the cutoff dependence of the masses and the root-mean-square (RMS) radius with the $S$-$D$ wave mixing analysis. [D] refers to the direct channel and [C] means the cross channel.}
\label{rr1}
\end{figure}

Since the $X(4630)$ has the decay channel $J/\psi\phi$ \cite{Aaij:2021ivw}, the $C$ parity of the $X(4630)$ is constrained as positive. We first study whether the newly observed $X(4630)$ can be assigned as the $D_s^{*}\bar D_{s1}$ molecular state with $J^{PC}=1^{-+}$.
For the $D_s^{*}\bar D_{s1}$ state with $J^{PC}=1^{-+}$, we study the bound properties by performing both the single channel and the $S$-$D$ wave mixing analysis. Here, we need to emphasize that the coupled channel effect to the $S$-wave $D_s^{*}\bar D_{s1}$ system is not obvious \cite{Wang:2020dya}. In Fig. \ref{rr1}, we present the bound state solutions for the $D_s^{*}\bar D_{s1}$ state with $J^{PC}=1^{-+}$ when considering the $S$-$D$ wave mixing effect. Here, the binding energy (mass) and RMS radius is  $-21$ MeV ($4626$ MeV) and 0.74 fm with $\Lambda=1.97$ GeV, this molecular state can correspond to the observed $X(4630)$ \cite{Aaij:2021ivw}. By comparing the numerical results of the single channel and $S$-$D$ wave mixing cases, we find that the $S$-$D$ wave mixing effect plays a minor role in generating the $D_s^{*}\bar D_{s1}$ bound state with $J^{PC}=1^{-+}$, and the dominant channel is the  $|{}^3\mathbb{S}_{1}\rangle$ with a probability around 99.84\%.

Indeed there exists relative phase of the potential involved in the $f_0(980)$ since the $f_0(980)$ cannot be included in the same framework as the $\eta$ and $\phi$. We further test the effect of this relative phase on our result as shown in Table~\ref{f0}. We may find that this relative phase cannot largely affect the results. It can be understood since the contribution from the $f_0(980)$ exchange is small and can be ignored compared with other exchange potentials from the $\eta$ and $\phi$.
\renewcommand\tabcolsep{0.20cm}
\renewcommand{\arraystretch}{1.50}
\begin{table}[!htbp]
\caption{The binding energy and root-mean square radius for the $D_s^{*}\bar D_{s1}$ state with $J^{PC}=1^{-+}$  with three different scenarios of the $f_0(980)$ exchange interaction. }\label{f0}
\begin{tabular}{c|ccc}
\toprule[1.0pt]
\toprule[1.0pt]
$f_0(980)$ exchange interaction &$E$ (MeV)&$\Lambda$ (GeV)&$r$ (fm)\\\midrule[1.0pt]
Attraction  &$-21.33$&1.969&0.74\\
None      &$-21.23$&1.979&0.74\\
Repulsion &$-21.14$&1.989&0.74\\\hline
\bottomrule[1.0pt]
\bottomrule[1.0pt]
\end{tabular}
\end{table}

By borrowing the experience of the former study of deuteron in the framework of one boson exchange, the cutoff $\Lambda$ is taken as around 1 GeV~\cite{Wang:2019nwt}. Usually, in realistic calculation \cite{Dong:2019ofp, Dong:2021juy, Zhu:2021lhd, He:2019csk, He:2019ify, He:2017mbh, Chen:2020kco, He:2014nxa}, setting the order of magnitude of the value of cutoff $\Lambda$ as $\mathcal{O}(1)$ is adopted. In this work, we find $\Lambda=1.97$ GeV when reproducing the central value of measured mass of the $X(4630)$. Since $\Lambda=1.97$ GeV is comparable with the requirement of $\mathcal{O}(1)$, we conclude that such cutoff value can be acceptable. Additionally, we need to indicate that the pion exchange is forbidden for the discussed $D^{\ast}_s\bar{D}_{s1}$ system for the $X(4630)$, we have to consider the $\eta$ and $\phi$ exchange contribution, which makes us to introduce large cutoff value when reproducing the mass of the $X(4630)$.

Under the framework of molecular state, the obtained RMS is not comparable with the hadronic molecular picture when reproducing the central value of mass of the $X(4630)$, where the binding energy reaches up to $-21.31$ MeV. We noticed a fact that there exist large error for the resonant parameter of the $X(4630)$. Thus, considering the error, the mass of $X(4630)$ can be $4644$ MeV, where the corresponding binding energy is $-3.31$ MeV. If reproducing such mass value, we find that the obtained RMS is 1.7 fm which is not in conflict with the hadronic molecular picture. Thus, we expect more precise measurement to clarify this point.

\begin{figure}[!htbp]
\includegraphics[width=4.20cm,keepaspectratio]{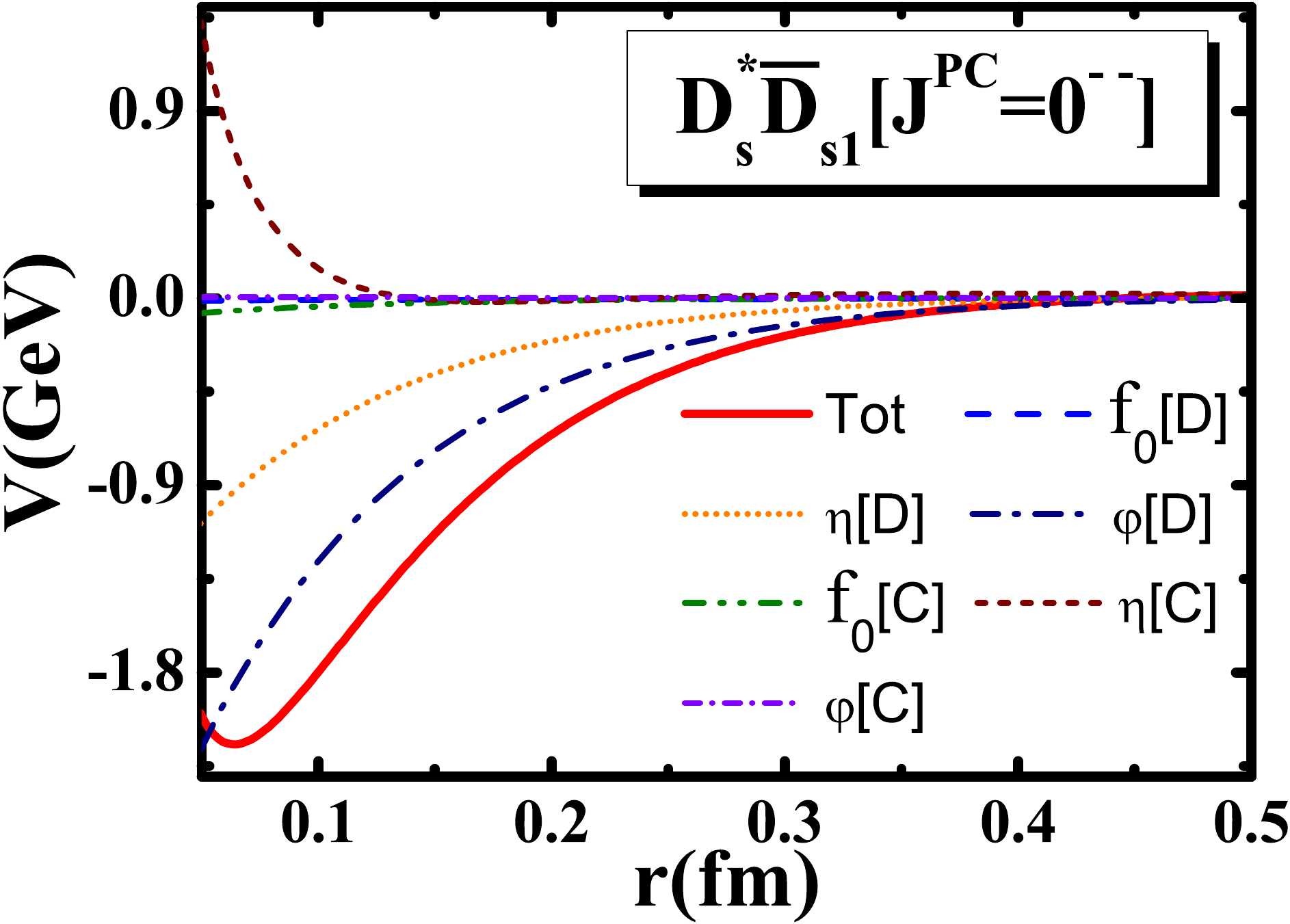}~\includegraphics[width=4.20cm,keepaspectratio]{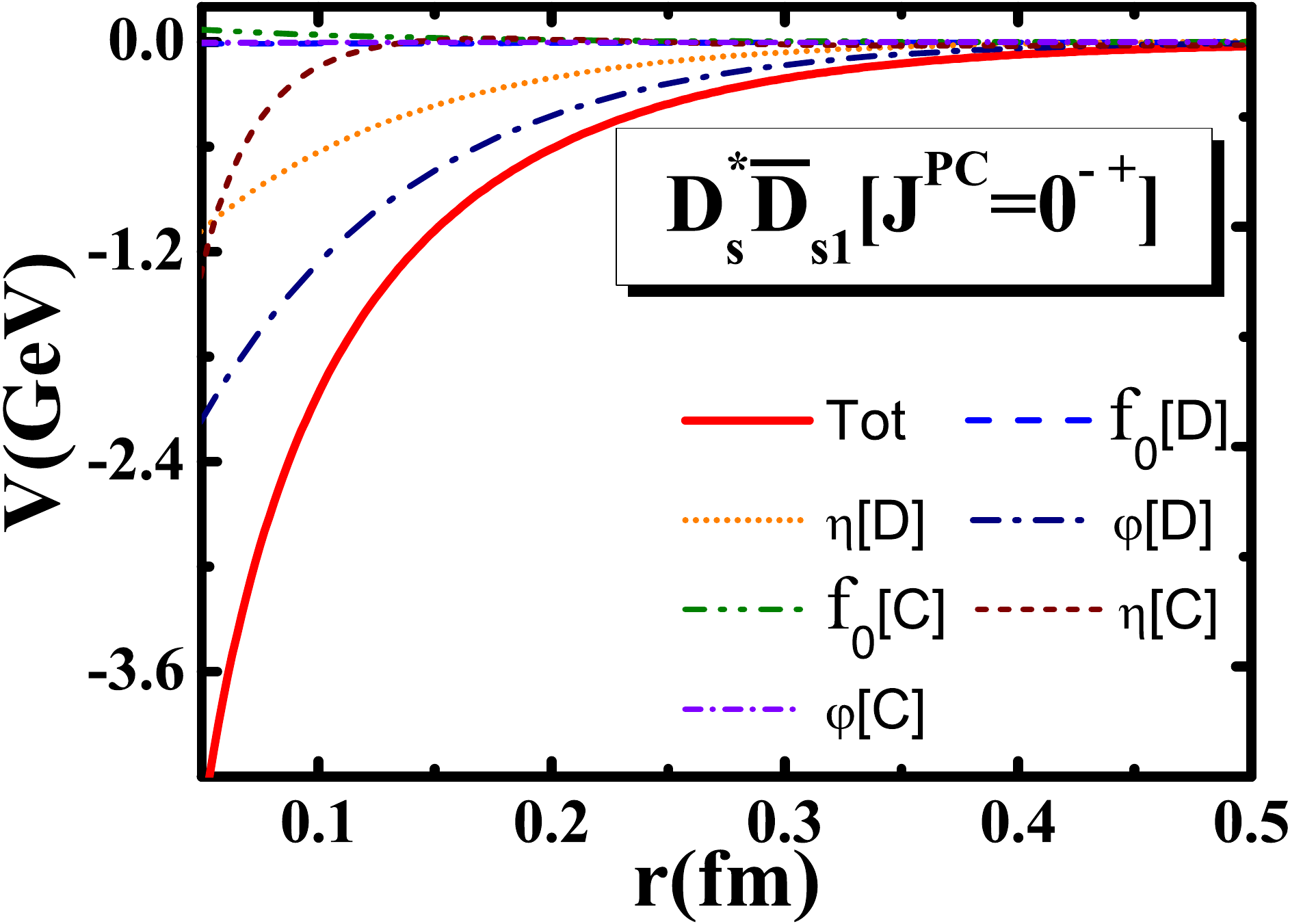}
\caption{The $r$ dependence of the OBE effective potentials for the $D_s^{*}\bar D_{s1}$ states with $J^{PC}$=$0^{--}$ and $0^{-+}$. Here, we take the cutoff $\Lambda=1.97$ GeV. [D] refers to the direct channel and [C] means the cross channel.}
\label{pp}
\end{figure}

In addition to explaining the $X(4630)$ as the $D_s^{*}\bar D_{s1}$ molecular state with $J^{PC}=1^{-+}$, we further predict the spin partner of the $X(4630)$. As shown in Fig. \ref{pp}, we present the $r$ dependence of the OBE potentials for the $D_s^{*}\bar D_{s1}$ states with $J^{PC}$=$0^{--}$ and $0^{-+}$.

\renewcommand\tabcolsep{0.09cm}
\renewcommand{\arraystretch}{1.50}
\begin{table}[!htbp]
\caption{Bound state solutions for the $S$-wave $D_s^{*}\bar D_{s1}$  system. The cutoff $\Lambda$, the binding energy $E$, and the RMS radius $r_{RMS}$ are in units of $ \rm{GeV}$, $\rm {MeV}$, and $\rm {fm}$, respectively.}\label{result}
\begin{tabular}{c|ccc|cccc}\toprule[1pt]\toprule[1pt]
\multicolumn{1}{c|}{Effect}&\multicolumn{3}{c|}{Single channel}&\multicolumn{4}{c}{$S$-$D$ wave mixing effect}\\\midrule[1.0pt]
$J^{PC}$&$\Lambda$ &$E$&$r_{\rm RMS}$&$\Lambda$ &$E$&$r_{\rm RMS}$&$P({}^1\mathbb{S}_{0}/{}^5\mathbb{D}_{0})$ \\
\cline{1-8}
\multirow{2}{*}{$0^{--}$}        &1.66&$-0.31$&4.49      &1.66&$-0.33$&4.41&\textbf{100.00}/$o(0)$\\
                                 &1.74&$-12.41$&0.87     &1.74&$-12.48$&0.87&\textbf{100.00}/$o(0)$\\
\cline{1-8}
\multirow{2}{*}{$0^{-+}$}        &1.56&$-0.63$&3.63     &1.55&$-0.28$&4.74&\textbf{99.96}/0.04\\
                             &1.63&$-11.49$&0.96      &1.63&$-11.96$&0.95&\textbf{99.88}/0.12\\\midrule[1.0pt]
$J^{PC}$&$\Lambda$ &$E$&$r_{\rm RMS}$&$\Lambda$ &$E$&$r_{\rm RMS}$&$P({}^3\mathbb{S}_{1}/{}^3\mathbb{D}_{1}/{}^5\mathbb{D}_{1})$ \\
\multirow{2}{*}{$1^{--}$}        &2.00&$-0.42$&4.16     &1.99&$-0.31$&4.53&\textbf{100.00}/$o(0)$/$o(0)$\\
                                 &2.14&$-12.13$&0.90    &2.13&$-12.07$&0.90&\textbf{99.96}/0.02/0.02\\
\bottomrule[1pt]\bottomrule[1pt]
\end{tabular}
\end{table}

In Table \ref{result}, we collect the corresponding bound state solutions. In our numerical analysis, we attempt to find the loosely bound solutions by varying the cutoff parameters $\Lambda$ from 1.00 to 2.50 $~{\rm GeV}$, and there may exist several possible hadronic molecular candidates, the $D_s^{*}\bar D_{s1}$ molecular states with $J^{PC}$=$0^{--}$, $0^{-+}$, and $1^{--}$. Thus, we have reason to believe that exploring these suggested hadronic molecules will be an interesting research issue, especially the $D_s^{*}\bar D_{s1}$ molecular states with $J^{PC}$=$0^{--}$ and $0^{-+}$, which can be taken as a crucial test of the molecular assignment of the $X(4630)$.

\section{ Hidden-charmed decay channels}
In addition to the mass spectrum, we also focus on the two-body hidden-charmed decay channels of the $D^{\ast}_{s}\bar{D}_{s1}$ system in the present study, that is the $(c\bar{s})(\bar{c}s)\to (c\bar{c})+(s\bar{s})$ process. It is the short-range interaction that makes the molecular states decay, and it is different from the long-range interaction that makes two hadrons bind together. In the study of the hidden-charmed decays, we adopt the quark-interchange model, which is introduced in Refs.~\cite{Hilbert:2007hc, Barnes:1991em, Barnes:1999hs, Barnes:2000hu, Wong:2001td, Wang:2019spc, Xiao:2019spy, Wang:2020prk, Wang:2021hql}. The one-gluon-exchange (OGE) potential $V_{ij}(q^2)$ is a good approximation to describe the interactions between the quarks, which is expressed as \cite{Wong:2001td}
\begin{eqnarray}
\begin{aligned}
V_{ij}(q^2)&=\frac{\lambda_i}{2}\cdot\frac{\lambda_j}{2}\left(\frac{4\pi\alpha_s}{q^2}+\frac{6\pi b}{q^4}-\frac{8\pi\alpha_s}{3m_im_j}e^{-{\frac{q^2}{4\sigma^2}}}\textbf{s}_i\cdot\textbf{s}_j\right),\\
\alpha_s(Q^2)&=\frac{12\pi}{(32-2n_f){\rm ln}(A+Q^2/B^2)},\label{e3}
\end{aligned}
\end{eqnarray}
where $Q^2$ is the square of the invariant masses of the interacting quarks, $\lambda_i$ represents the color Gell-Mann matrix, $m_i$ is the quark mass, and $\textbf{s}_i$ represents the spin operator of the interacting quarks. The adopted parameters related to the OGE potential are $m_s=0.575~\rm{GeV}$, $m_c=1.776~\rm{GeV}$, $b=0.180~\rm{GeV}$, $\sigma=0.897~\rm{GeV}$, $A=10$, and $B=0.310~\rm{GeV}$ \cite{Wong:2001td}.

The meson wave function can be written as
\begin{eqnarray}
\psi=\omega_{\rm{color}}\chi_{\rm{flavor}}\chi_{\rm{spin}}\phi(\textbf{p}).\label{mesonwavefunction}
\end{eqnarray}
In this letter, we take the single Gaussian function to approximate the momentum space wave function of the meson
\begin{eqnarray}
\phi(\textbf{p}_{\rm{rel}})=&2^{\frac{l}{2}}\pi^{-\frac34}\beta^{-\frac{3}{2}-l}p^l
\sqrt{\frac{4\pi}{(2l+1)!!}} {\rm Y}_{lm}(\hat{\Omega})e^{-\frac{\textbf{p}_{\rm{rel}}^2}{2\beta^2}}. \label{e7}
\end{eqnarray}
Here, $\beta$ denotes the oscillating parameter of the Guassian function, $\textbf{p}_{\rm{rel}}=(m_{\bar{q}}\textbf{p}_q+m_q\textbf{p}_{\bar{q}})/(m_q+m_{\bar{q}})$ is the relative momentum with $m_q$ ($m_{\bar{q}}$) and $\textbf{p}_q$ ($\textbf{p}_{\bar{q}}$) being the masses and momenta of the quarks (anti-quarks) in the meson, ${\rm Y}_{lm}(\hat{\Omega})$ is the orbital angular function. The parameters $\beta$ are fitted by the mass spectrum of the mesons, and the numerical values are $\beta_{D^{\ast}_s}=0.440$, $\beta_{D_{s1}}=0.385$, $\beta_{\phi}=0.370$, $\beta_{\eta^{(\prime)}}=0.465$, $\beta_{\eta_{c}}=0.618$, $\beta_{\eta_{c}(2S)}=0.471$, $\beta_{J/\psi}=0.595$, and $\beta_{\chi_{cJ}(1P)}=0.500$ in units of GeV \cite{Zyla:2020zbs}.

The wave function for the molecular state composed of two mesons $A$ and $B$ in momentum space is similar to the meson wave function, and the corresponding $\beta$ parameter can be written as $\beta=\sqrt{3M_AM_B(M_A+M_B-M)/(M_A+M_B)}$. Here, $M_A$, $M_B$, and $M$ are the masses of the meson $A$, the meson $B$, and the molecular state~\cite{Guo:2017jvc, Chen:2017xat, Weinberg:1962hj, Weinberg:1963zza}.

The two-body strong decay widths of these discussed molecular candidates can be calculated by
\begin{eqnarray}
\begin{aligned}
\Gamma=&\frac{|\textbf{P}_C|}{32\pi^2M^2(2J+1)}\int d\Omega|\mathscr{M}_{fi}|^2,\label{e3}
\end{aligned}
\end{eqnarray}
where $|\textbf{P}_C|$ represents the three-momentum of the final state in the center-of-mass reference frame. According to the quark potential and the wave function, the scattering matrix $\mathscr{M}_{fi}$ can be given as the product of the following factors, i.e.,
\begin{eqnarray}
\mathscr M_{fi}=K I_{\rm{color}}I_{\rm{flavor}}I_{\rm{spin}}I_{\rm{space}},\label{factor}
\end{eqnarray}
where $K=(2\pi)^{\frac{3}{2}}\sqrt{2M}\sqrt{2E_C}\sqrt{2E_D}$, and $E_C$ and $E_D$ stand for the energies of the mesons in the final states. The specific calculation can be referred to Refs.~\cite{Wong:2001td, Wang:2019spc, Xiao:2019spy, Wang:2020prk}.

For the $D^{\ast}_s\bar{D}_{s1}$ state with $J^{PC}=1^{-+}$, it may decay into the $\eta_c\eta$, $\eta_c\eta^{\prime}$, $J/\psi\phi$, $\eta_c(2S)\eta$, $\eta_c(2S)\eta^{\prime}$, $\chi_{c1}(1P)\eta$, $\chi_{c1}(1P)\eta^{\prime}$, and so on. By performing numerical calculation, we have
\begin{eqnarray}
&\Gamma_{J/\psi\phi}:\Gamma_{\eta_c(2S)\eta^{\prime}}:\Gamma_{\eta_c(2S)\eta}:\Gamma_{\chi_{c1}(1P)\eta^{\prime}}:\Gamma_{\chi_{c1}(1P)\eta}\nonumber\\
&=1:0.56:0.43:0.09:0.04.\label{ratio}
\end{eqnarray}
Other hidden-charmed decay widths are smaller than $1\%$ of that for the $J/\psi\phi$ decay channel, and thus we do not show here.

\begin{figure}[!tbp]
\centering
\begin{tabular}{c}
\includegraphics[scale=0.38]{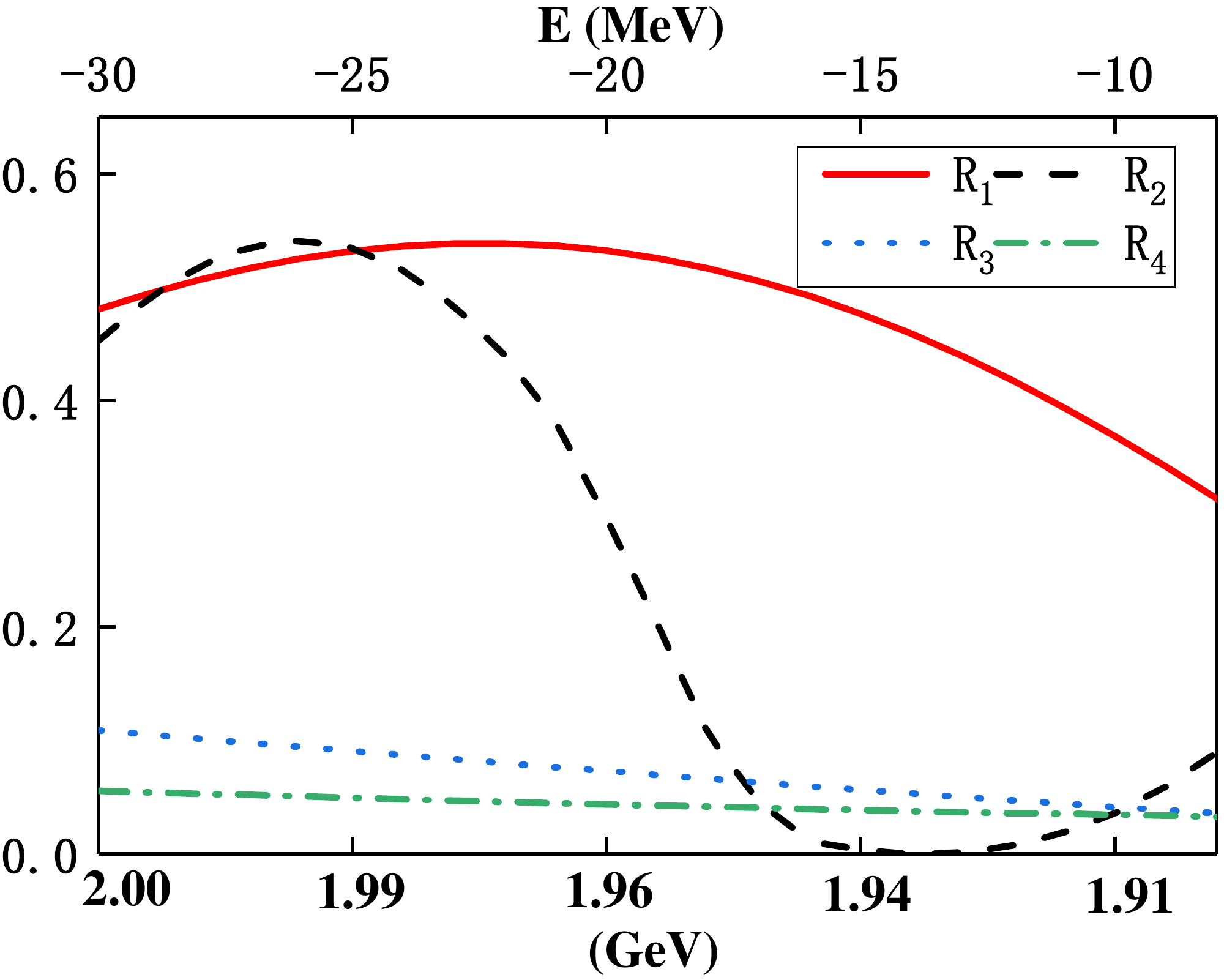}\\
\end{tabular}
\caption{The dependence of the decay ratios on cutoff $\Lambda$ and binding energy $E$ for the $D^{\ast}_{s}\bar{D}_{s1}$ the state with $J^{PC}=1^{-+}$. Here, we defined $R_1=\Gamma_{\eta_c(2S)\eta^{\prime}}/\Gamma_{J/\psi\phi}$, $R_2=\Gamma_{\eta_c(2S)\eta}/\Gamma_{J/\psi\phi}$, $R_3=\Gamma_{\chi_{c1}(1P)\eta^{\prime}}/\Gamma_{J/\psi\phi}$, and $R_4=\Gamma_{\chi_{c1}(1P)\eta}/\Gamma_{J/\psi\phi}$. }
\label{relative partial widths}
\end{figure}

Considering the uncertainty of the mass of the $X(4630)$, we present the dependence of the decay ratios on cutoff $\Lambda$ and binding energy $E$ in Fig.~\ref{relative partial widths}. We find that the $J/\psi\phi$ mode is still significant decay compared with other hidden-charm decay channels. Thus, this result still explain why the $X(4630)$ was firstly observed in its $J/\psi\phi$ decay mode. And, most of charmoniumlike $XYZ$ states were observed in their hidden-charm decay channels like $J/\psi$ plus some light mesons. Indeed, from experimental side, it can be understood that the $J/\psi$ is more easier to be constructed than other charmonia.

For the $D^{\ast}_s\bar{D}_{s1}$ bound state with $J^{PC}=1^{-+}$, it can decay into the $\eta_{c}\eta^{(\prime)}$ and $J/\psi\phi$ channels through the $P$-wave interaction, but the widths of the $\eta_{c}\eta^{(\prime)}$ channels are much smaller than the $J/\psi\phi$ channel despite the larger phase spaces. As shown in Eq.~(\ref{e3}), the decay width depends on the relative momentum $|\textbf{P}_C|$ in the final states and the square of transition amplitude $|\mathscr{M}_{fi}|^2$. The larger $|\textbf{P}_C|$ may lead to the smaller transition amplitude $\mathscr{M}_{fi}$~\cite{Wang:2019spc,Wang:2021aql}, and thus the decay width may become smaller.

\begin{figure}[!tbp]
\centering
\includegraphics[scale=0.59]{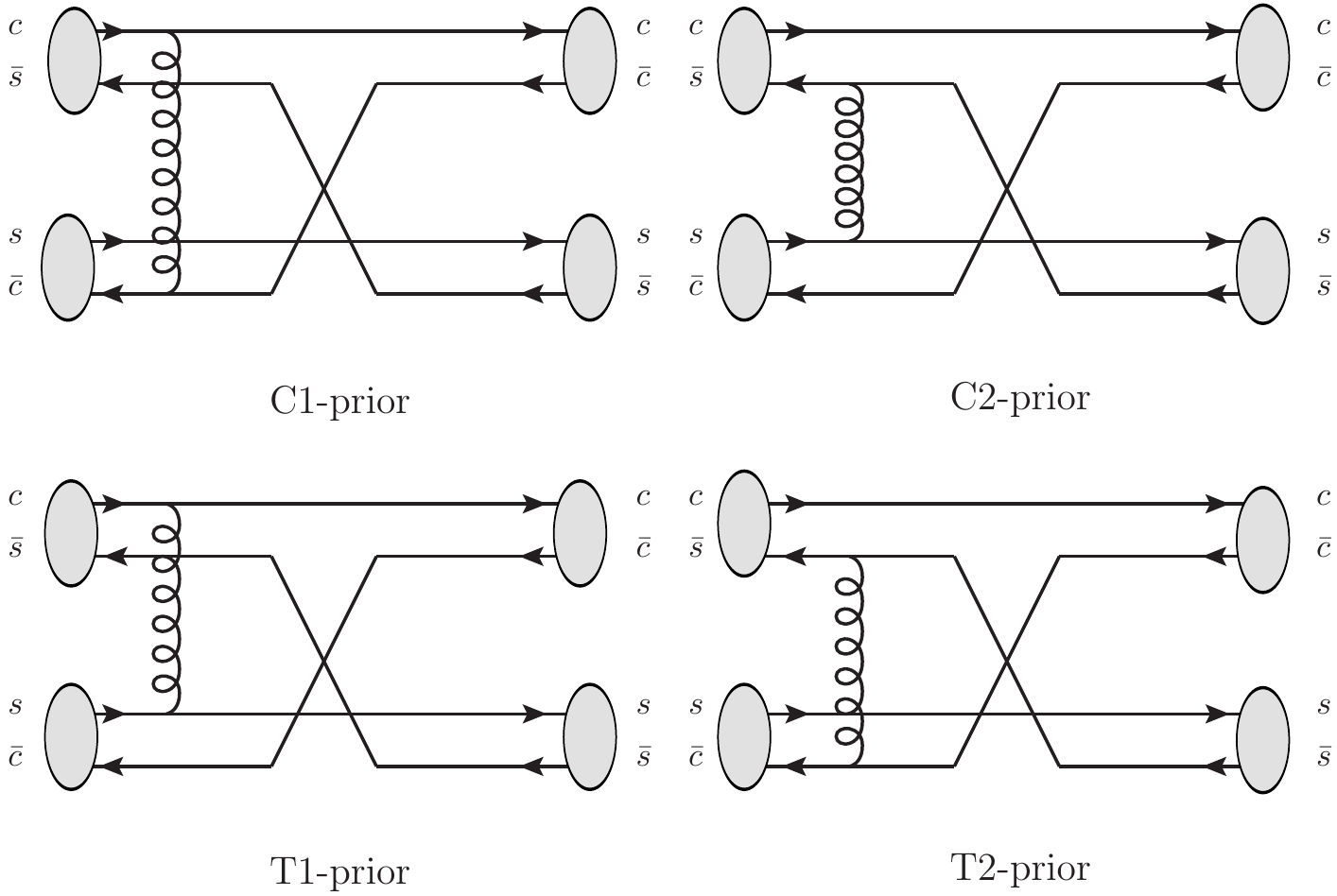}
\caption{Quark-interchange diagrams for the scattering process $A(c\bar{s})+B(\bar{c}s)\to C(c\bar{c})+D(s\bar{s})$ in the molecular picture \cite{Wang:2021aql}. The curve line denotes the (anti-)quark-(anti-)quark interactions.}\label{interchangediagrams}
\end{figure}

The transition amplitude $\mathscr{M}_{fi}$ for the scattering process $c\bar{s}+\bar{c}s\to c\bar{c}+s\bar{s}$ receives the contributions from the four diagrams of Fig.~\ref{interchangediagrams} within the quark-interchange model \cite{Wang:2021aql}. For the $D^{\ast}_s\bar{D}_{s1}$ bound state with $J^{PC}=1^{-+}$, the signs of the Feynman amplitudes from the four quark-interchange diagrams are different for the $\eta_{c}\eta^{(\prime)}$ decay channel, and the contributions largely cancel among them, which leads to the suppression of the decay width. The decay widths may be very different for different spin structures of the $J/\psi \phi$ and $\eta_c \eta^{(\prime)}$ decay channels, and one can find similar situations in Ref. \cite{Xiao:2019spy}. 

Furthermore, the spin factor $I_{\rm{spin}}$ in Eq.~(\ref{factor}) for the $\eta_{c}\eta^{(\prime)}$ channel is smaller than that for the $J/\psi\phi$ channel. $|I_{\rm{spin}}|_{J/\psi\phi}^2:|I_{\rm{spin}}|_{\eta_{c}\eta^{(\prime)}}^2$ is 2 or 3 for the diagram in Fig. ~\ref{interchangediagrams}, which also partly contributes to the suppression.

The similar decay suppressions are also noticed with other approaches ~\cite{Lin:2017mtz, Chen:2017xat, Lin:2018kcc, Lin:2018nqd, Shen:2019evi, Lin:2019qiv, Lin:2019tex, Dong:2019ofp, Dong:2020rgs, Chen:2017abq, Xiao:2016mho, Xiao:2019mvs, Wu:2018xaa,Wang:2018pwi,Wang:2021aql}. However, we need to point out that the decays of the molecular states cannot always be predicted very reliably and are still deserved to study furthermore in future.

\begin{figure}[!tbp]
\centering
\includegraphics[scale=0.48]{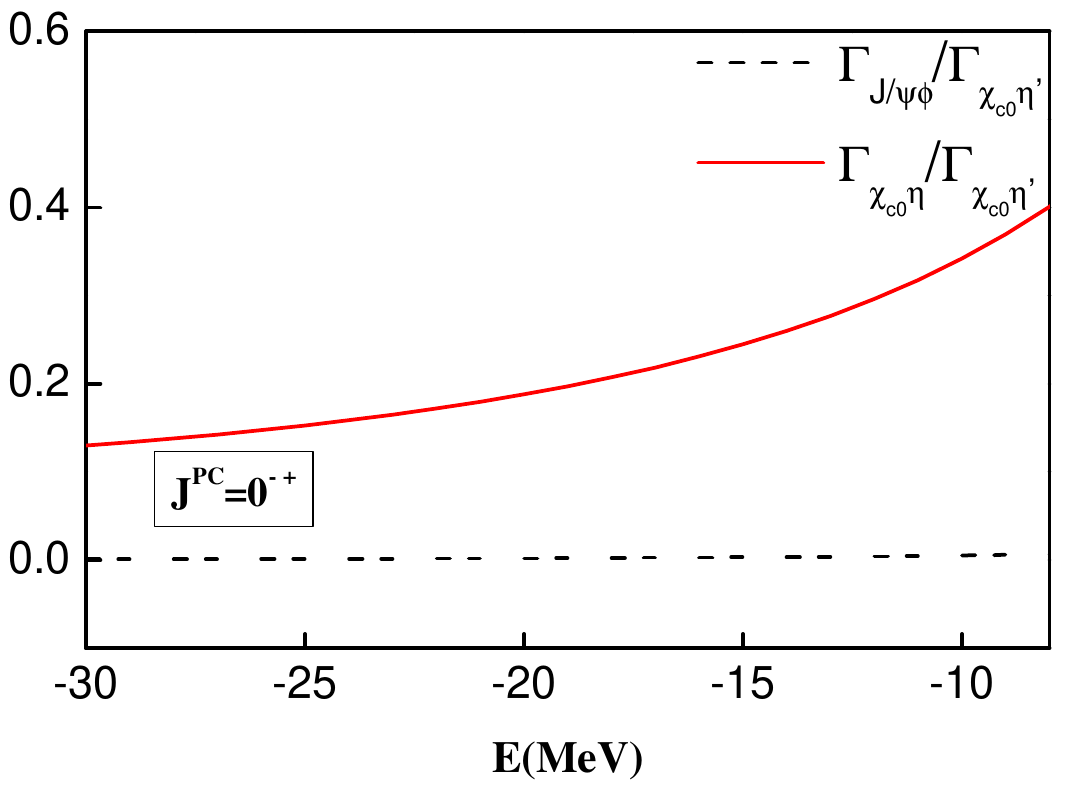}
\includegraphics[scale=0.48]{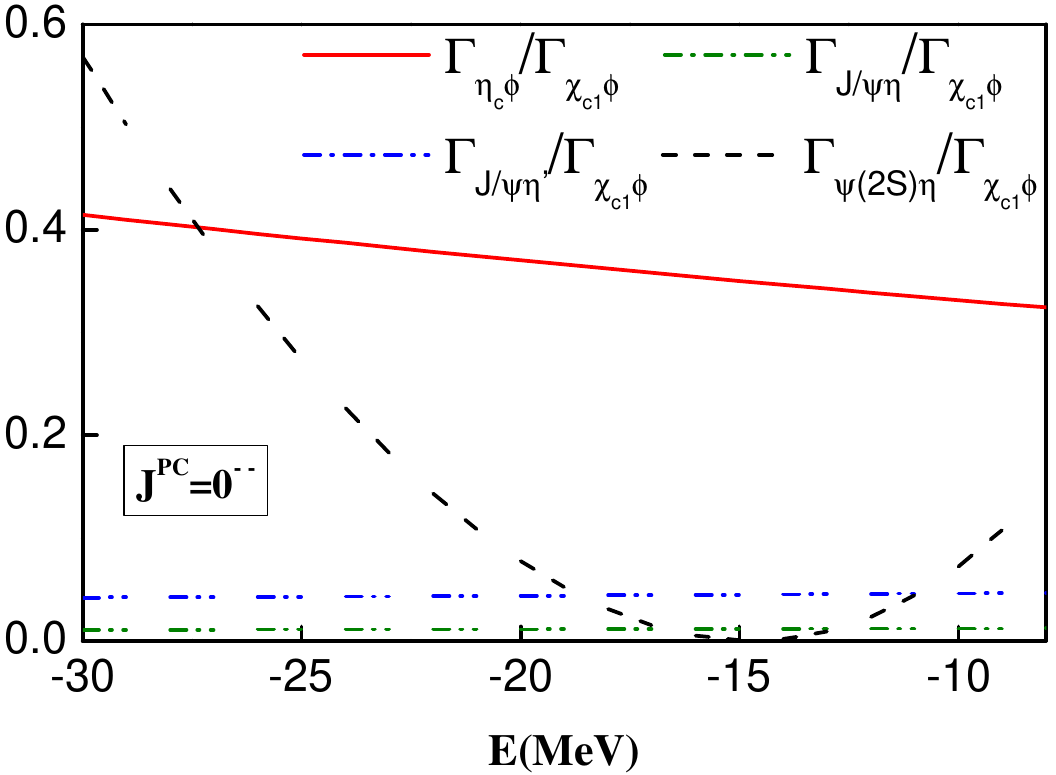}
\includegraphics[scale=0.48]{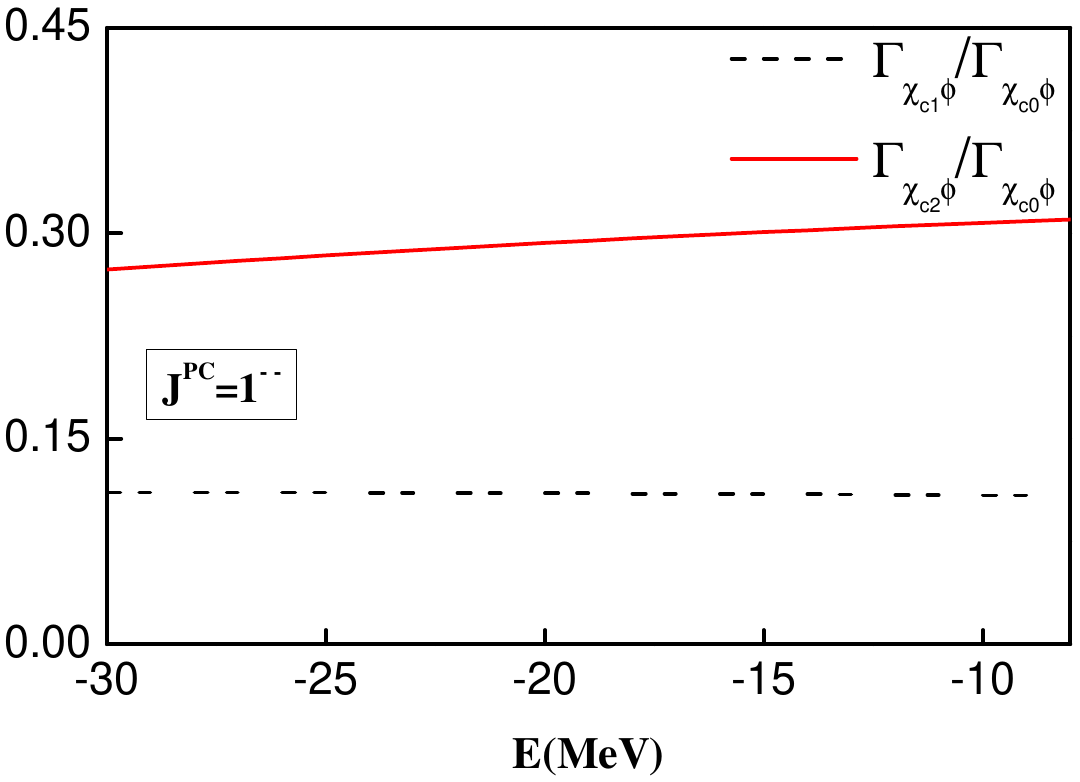}
\caption{The binding energy dependence of the decay width ratios for the $D^{\ast}_s\bar{D}_{s1}$ molecular candidates with $J^{PC}=0^{-+},0^{--},1^{--}$, respectively.}
\label{partner}
\end{figure}

In Fig.~\ref{partner}, we present the binding energy dependence of the decay width ratios for the partners of the $X(4630)$.  From the figure, we can get the decay properties of the $X(4630)$ partners.
\begin{itemize}
  \item For the $D^{\ast}_s\bar{D}_{s1}$ molecular candidate with $J^{PC}=0^{-+}$, there exists the $J/\psi\phi$, $\chi_{c0}(1P)\eta$, and $\chi_{c0}(1P)\eta^{\prime}$ decay modes, and it is much easier to be detected in the $\chi_{c0}(1P)\eta^{\prime}$ channel.
  \item For the $D^{\ast}_s\bar{D}_{s1}$ molecular candidate with $J^{PC}=0^{--}$, it can decay into the $\eta_c\phi$, $J/\psi\eta$, $J/\psi\eta^{\prime}$, $\psi(2S)\eta$, and $\chi_{c1}(1P)\phi$ channels, and it couples strongly with the $\chi_{c1}(1P)\phi$ channel.
  \item The $D^{\ast}_s\bar{D}_{s1}$ molecular candidate with $J^{PC}=1^{--}$ allows decay channels including the $\chi_{c0}(1P)\phi$, $\chi_{c1}(1P)\phi$, and $\chi_{c2}(1P)\phi$, and it prefers to decay into the $\chi_{c0}(1P)\phi$ channel.
\end{itemize}

\section{Summary}
There is no end to the exploration of the matter world. At present, investigation of hadron spectroscopy is bringing our surprises. As announced by LHC on March 3 2021 \cite{lhcb}, 59 new hadrons were discovered in the past decade. This situation shows that it is an active research field full of opportunities.

Inspired by the recent observation of new resonance $X(4630)$ existing in the $J/\psi\phi$ invariant mass spectrum of $B^+\to J/\psi\phi K^+$ \cite{Aaij:2021ivw}, we notice the peculiar feature of the $X(4630)$, which has exotic $J^{PC}=1^{-+}$ quantum number different from that of conventional meson and is near the $D_s^{*} D_{s1}(2536)$ threshold. In this letter, we propose that the newly observed $X(4630)$ resonance is a good candidate of charmoniumlike molecule. By OBE mechanism, the mass of $D_s^{*}\bar D_{s1}(2536)$ charmoniumlike molecule with $J^{PC}=1^{-+}$ is calculated to be consistent with that of the $X(4630)$, which is the first proof of supporting the $X(4630)$ as a charmoniumlike molecule. We also carry out further study of the corresponding two-body hidden-charmed decays of the $X(4630)$, and find the dominant role of the $J/\psi\phi$. This decay behavior naturally lets us understand why the $X(4630)$ was firstly discovered by analyzing its $J/\psi\phi$ decay channel in LHCb \cite{Aaij:2021ivw}.

The success of decoding the $X(4630)$ as a charmoniumlike molecule may enforce our ambition in constructing charmoniumlike molecule zoo. In this letter, we further predict the spin partner of the $X(4630)$. Searching for the spin partner of the $X(4630)$ will become an intriguing research issue at LHC.

Facing the present prosperous situation of new hadrons, we have strong confidence to believe that more states will be filled in the zoo of charmoniumlike molecule  in future with the joint effort from both theorist and experimentalist.

\section*{ACKNOWLEDGMENTS}
This work is supported by the China National Funds for Distinguished Young Scientists under Grant No. 11825503, National Key Research and Development Program of China under Contract No. 2020YFA0406400, the 111 Project under Grant No. B20063, and the National Natural Science Foundation of China under Grant No. 12047501. This project is also supported by the National Natural Science Foundation of China under Grants No. 12175091, and 11965016, CAS Interdisciplinary Innovation Team, and the Fundamental Research Funds for the Central Universities under Grants No. lzujbky-2021-sp24.

\end{document}